\documentclass[10pt,a4paper]{article}
\usepackage[utf8]{inputenc}
\usepackage{authblk}
\usepackage{amsmath}
\usepackage{amsfonts}
\usepackage{amssymb}
\usepackage{graphicx}
\usepackage{lmodern}
\usepackage{times}
\usepackage{caption}
\usepackage{mathptmx}
\usepackage{enumitem}
\title{NOvA Recent Results with Neutrino+Antineutrino Data}
\author{Tomas Nosek\footnote{e-mail: tomas.nosek@mff.cuni.cz}~(on behalf of the NOvA Collaboration)}
\affil{\small\textit{Charles University, Faculty of Mathematics and Physics, Institute of Particle and Nuclear Physics, V Holesovickach 2, 180 00 Prague, Czech Republic}}
\date{}

\begin{document}
\maketitle
\centerline{\begin{minipage}{.8\textwidth}
\small
\paragraph*{Abstract.} NOvA is a long-baseline neutrino oscillation experiment using Fermilab's 700 kW NuMI muon neutrino beam. Two functionally identical scintillator detectors are placed off the beam axis, separated by 810 km oscillation baseline. Both detectors have high active material fractions and are finely segmented allowing for precise identification and analysis of neutrino interactions. By observing both the disappearance of muon (anti)neutrinos and appearance of electron (anti)neutrinos in the beam, NOvA can impose constraints on the yet undetermined parameters of neutrino oscillation phenomenon, such as the neutrino mass ordering, CP violation and the octant of the large mixing angle. NOvA also studies neutral-current neutrino interactions, thus extending its scope beyond the standard three-flavor paradigm. This paper presents the latest NOvA results with the complete neutrino data sample up to date and first antineutrino data collected since February 2017.
\end{minipage}}

\section{Introduction}
NOvA is a long-baseline neutrino oscillation experiment designed to make measurements of muon neutrinos ($\nu_\mu$) disappearance and electron neutrinos ($\nu_e$) appearance in Fermilab's NuMI (Neutrinos at the Main Injector) beam. Well tuned for the first oscillation maximum around neutrino energy of 2 GeV over 810~km baseline, the experiment studies primarily four channels of oscillations: $\nu_\mu\rightarrow\nu_\mu$ or $\nu_\mu \rightarrow \nu_e$ and $\bar{\nu}_\mu\rightarrow\bar{\nu}_\mu$ or $\bar{\nu}_\mu \rightarrow \bar{\nu}_e$. They allow to address several concerns of neutrino oscillations:
\begin{enumerate}[noitemsep, nolistsep]
\item mass ordering, i.e.~normal (NH) or inverted hierarchy (IH) of neutrino mass eigenstates,
\item direct CP violation ($\delta_{\mathrm{CP}}$ phase) and
\item precise determination of $\theta_{23}$ and $\Delta m_{32}^2$ neutrino mixing parameters.
\end{enumerate}
\par This paper reports the 2018 NOvA combined analysis of $8.85 \times 10^{20}$~POT (protons on target) neutrino data collected from Feb 2018 to Feb 2017 and $6.91 \times 10^{20}$ POT antineutrino data collected from Feb 2017 to Apr 2018. Neutrino oscillations parametrization, fits, predictions and interpretation of the results were done within the standard oscillation model of 3 active neutrino flavors of electron, muon and tau neutrino ($\nu_\tau$)~\cite{PDG}.
\section{The NOvA Experiment}
The experiment consists of two large functionally identical detectors sitting\linebreak 14.6~mrad off the beam axis 810 km apart. This off-axis configuration reduces uncertainty on energy of incoming neutrinos and suppresses the higher-energy neutrinos background producing neutral current interactions (NC) misidentified as $\nu_e$ charged current (CC). On the other hand, it also results in a lower intensity than in the on-axis region, mitigated by the size of the detectors and beam power upgrades.
\par The detectors are finely grained and high active ($\sim65\%$ active mass) liquid scintillator tracking calorimeters, which allow for precise analysis of investigated neutrino interactions events. They are designed to be as similar as possible aside from size: the far detector (FD) is 14 kt and on surface located in Ash River, Minnesota, the near detector (ND) is located underground in Fermilab, close enough to the neutrinos source to see a far greater flux with only 0.3~kt of mass. Both are constructed out of extruded PVC cells (3.9$\times$6.6 cm in cross section and 15.5/3.8~m in length for FD/ND) filled with scintillator and equipped with a wavelength shifting fiber connected to avalanche photodiode (APD). They collect light produced by charged particles subsequently amplified and digitized by APDs. The cells alternate in horizontal and vertical orientation to allow for a stereo readout. More information on detectors can be found in Ref.~\cite{TechNote}.
\par NuMI beam is created following the decay of charged pions and kaons produced by 120 GeV protons hitting a carbon target. These parent mesons are focused by two magnetic horns towards the NOvA detectors and decay in flight through the chain $K^+, \pi^+\rightarrow \mu^+ + \nu_\mu$, with the muon then decaying as $\mu^+ \rightarrow e^+ + \nu_e + \bar{\nu}_\mu$. By switching the polarity of the horns, opposite charge sign particles can be focused, thus effectively selecting an antineutrino beam. The resulting composition in range 1-5 GeV at ND is of 96\% $\nu_\mu$, 3\% $\bar{\nu}_\mu$ and 1\% $\nu_e+\bar{\nu}_e$ in case of neutrino beam and 83\% $\bar{\nu}_\mu$, 16\% $\nu_\mu$ and 1\% $\bar{\nu}_e + \nu_e$ in case of antineutrino~beam.

\par To identify and classify neutrino interactions NOvA uses a method based on image recognition techniques known as Convolutional Visual Network (CVN), see Ref.~\cite{CVNpaper}. CVN treats every interaction in the detector as an image with cells being pixels and collected charge being their color. When trained with simulated events and cosmic data, CVN can extract abstract topological features of neutrino-like interactions with convolutional filters (feature maps \cite{CVNpaper}). With an input of calibrated 2D pixelmap (two views of horizontal and vertical event projections), the output is a set of normalized classification scores ranging over hypotheses of beam neutrinos event ($\nu_\mu$ CC, $\nu_e$ CC, $\nu_\tau$ CC and NC) or cosmics. CVN has been used together with additional supporting PIDs: seperate $\nu_e$ and $\nu_\mu$ cosmic rejection boosted decision trees (BDT) and muon track identification in $\nu_\mu$ events.
\par
\begin{figure}
	\includegraphics[width=\linewidth]{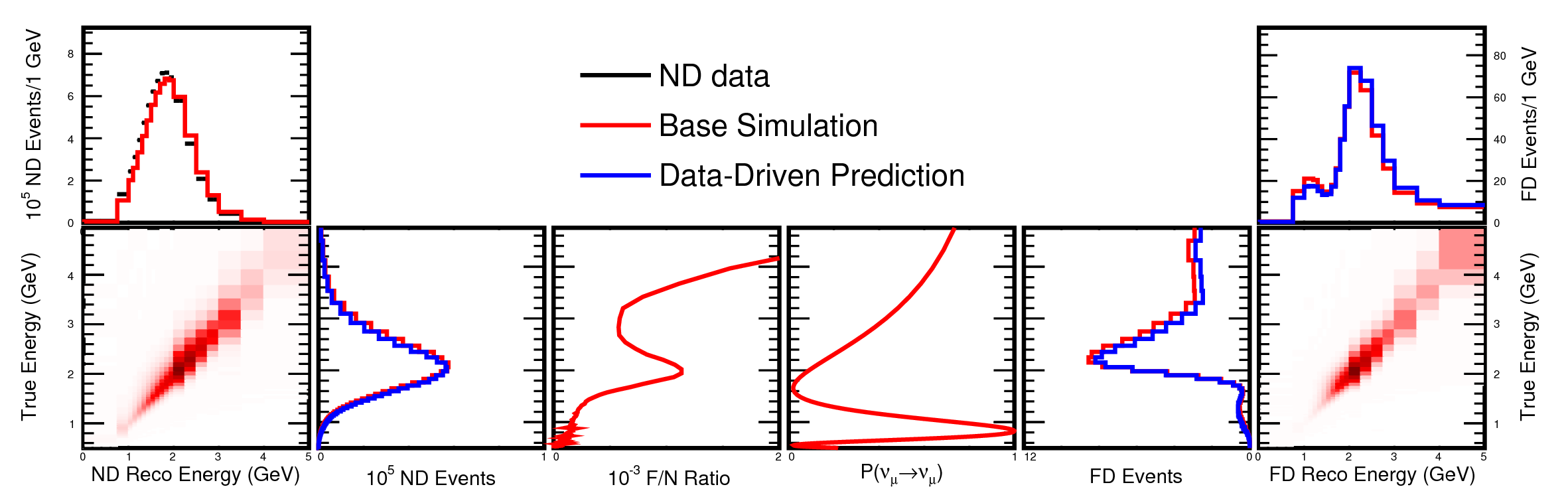}
	\caption{An illustration of NOvA's F/N technique. From left to right: reconstructed to true $\nu_\mu$ energy translation, F/N ratio, $\nu_\mu \rightarrow \nu_\mu$ oscillation probability, true to reconstructed $\nu_\mu$ energy restoration. Base simulation in red, ND data-driven corrected prediction in blue.}
	\label{FigExtrap}
\end{figure}

\begin{figure}
	\begin{center}
		\includegraphics[width=.496\linewidth]{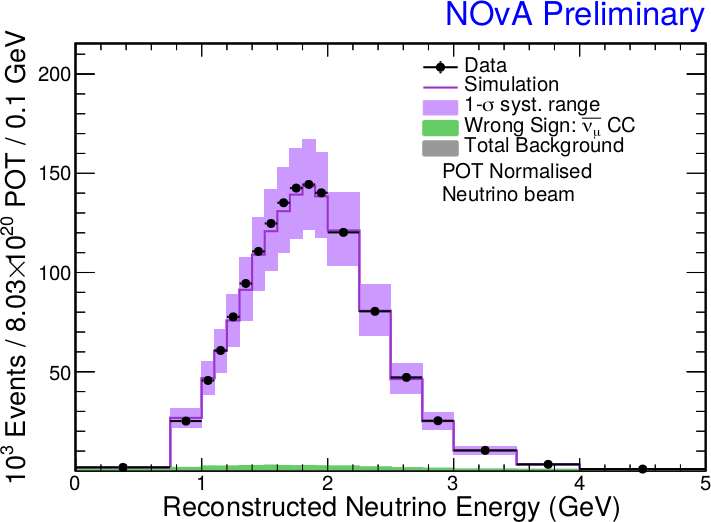}
		\includegraphics[width=.496\linewidth]{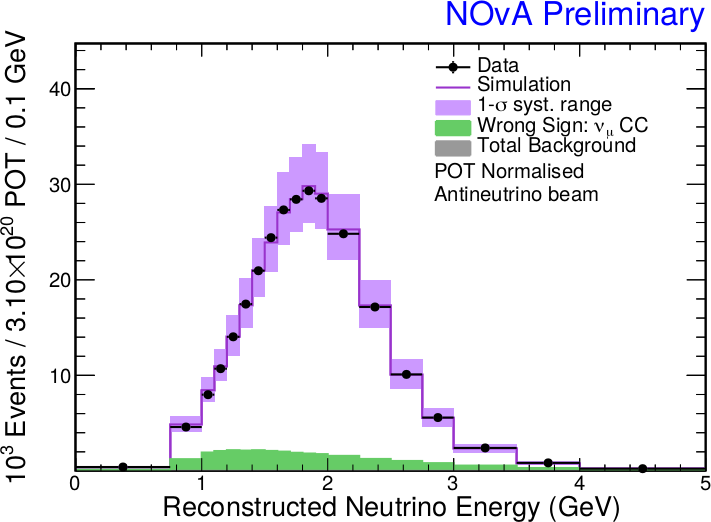}
	\end{center}
	\caption{ND selected $\nu_\mu$ (left) and $\bar{\nu}_\mu$ (right) reconstructed energy in data (black dots) and simulation (purple). Each bin is normalized by its width.}
	\label{FigNDnumu}
\end{figure}

 NOvA's two identical detectors design enables to employ data-driven predictions of FD observations. ND neutrinos spectra are considered the effective unoscillated source of oscillated neutrinos measured in FD. FD $\nu_\mu$ and $\nu_e$ signal is predicted using ND $\nu_\mu$, whereas FD $\nu_e$ beam background is constrained using ND $\nu_e$ sample. This Far/Near (F/N) technique includes several steps (Fig.~\ref{FigExtrap}). Firstly, the reconstructed neutrino energy spectrum is unfolded to true energy using a simulated migration matrix. Secondly, F/N ratio accounting for geometry, beam divergence and detector acceptance is applied to create unoscillated FD prediction. Then the FD spectrum is weighted by the oscillation probability for a given set of oscillation parameters. Finally, the true energy is smeared back again to the reconstructed energy via the migration matrix. As a reward, F/N technique significantly reduces both neutrino flux and cross section systematic uncertainties. ND reconstructed energy spectra of $\nu_\mu$ and $\bar{\nu}_\mu$ (the source of FD $\nu_\mu$ and $\nu_e$ signals) can be found in Fig.~\ref{FigNDnumu}.  

\section{Muon Neutrino and Antineutrino Disappearance}
Muon neutrino disappearance channel is primarily sensitive to $\vert \Delta m_{32}^2 \vert$ and $\sin^2 2\theta_{23}$ and the precision with which they can be measured depends on the $\nu_\mu$ energy resolution. Energy of $\nu_\mu$ is reconstructed as a sum of the energy of a muon, estimated from the range of muon track, and remaining hadronic energy. To get best effective use of the energy resolution, the data binning is optimized in two ways. First, the energy binning has finer bins near the disappearance maximum and coarser bins elsewhere. Second, the events in each energy bin are further divided into four populations, or ``quartiles'', of varying reconstructed hadronic energy fraction, which correspond to different $\nu_\mu$ energy resolution. The divisions are chosen such that the quartiles are of equal size in the unoscillated FD simulation. The $\nu_\mu$ ($\bar{\nu}_\mu$) energy resolution is estimated to be 5.8\% (5.5\%), 7.8\% (6.8\%), 9.9\% (8.3\%) and 11.7\% (10.8\%) for each quartile, ordered from lower to higher hadronic energy fraction. F/N technique is applied separately in quartiles, which has the additional advantage of isolating most of the cosmic and beam NC background events along with events of worst energy resolution (4$^{th}$ quartile).
\par
\begin{figure}
	\begin{center}
	\includegraphics[width=.47\linewidth]{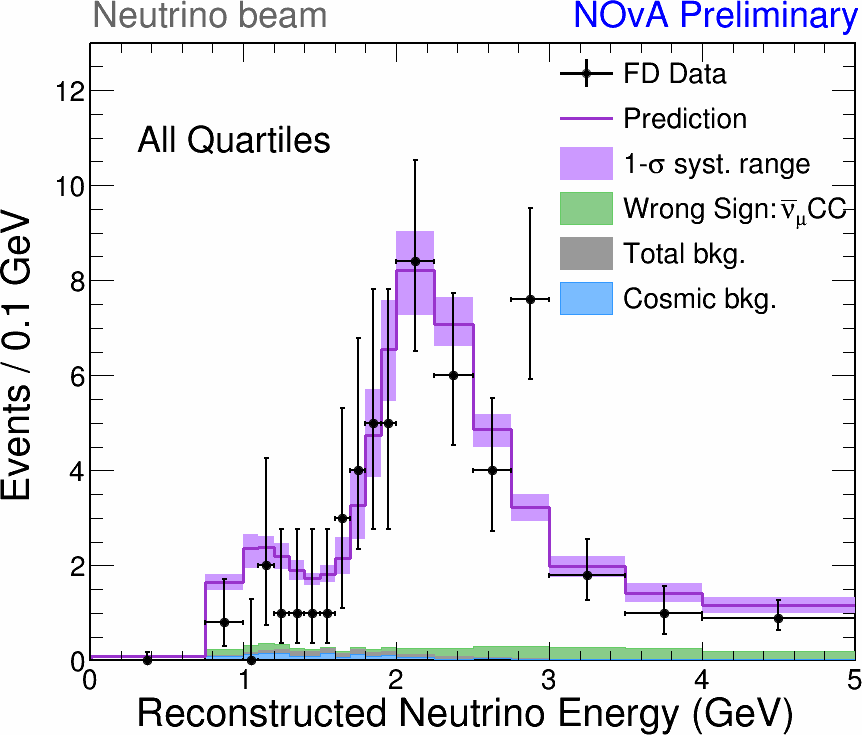}\hspace*{.04\linewidth}
	\includegraphics[width=.47\linewidth]{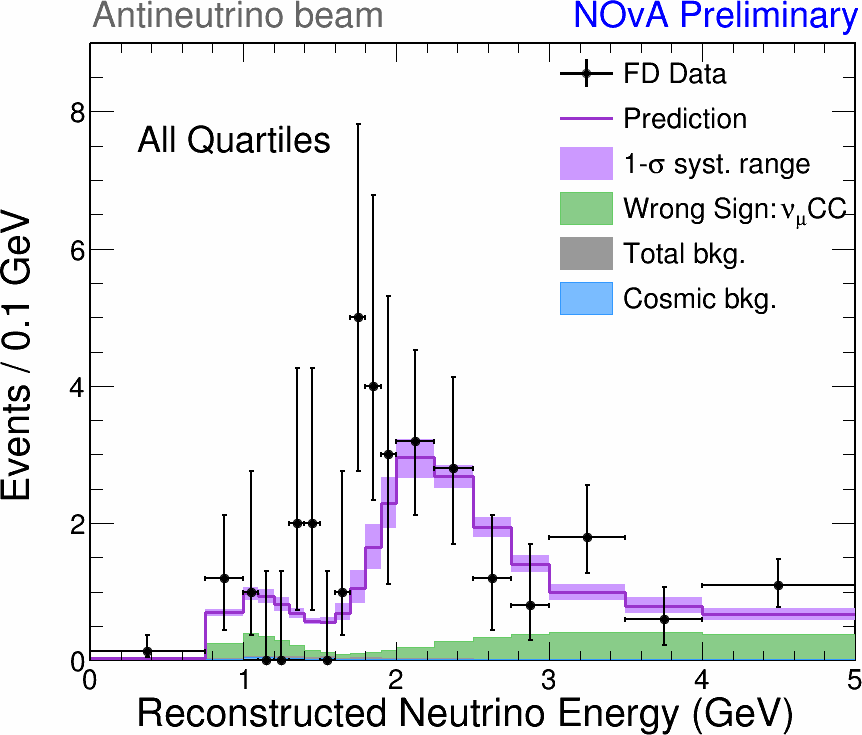}
	\end{center}
	\caption{FD data (black dots) selected $\nu_\mu$ (left) and $\bar{\nu}_\mu$ (right) candidates reconstructed energy compared to the best fit prediction (purple line) with 1$\sigma$ systematics uncertainty range. Summed over all quartiles of hadronic energy fraction.}
	\label{FigFDnumu}
\end{figure}

 There are 113 (65) $\nu_\mu$ ($\bar{\nu}_\mu$) CC candidates observed in FD, whereas with no oscillations, the projected ND flux would result in 730$^{+38}_{-49}$(syst.)$\pm27$(stat.) (266$^{+12}_{-14}\pm16$) events. The total estimated background is 3.76 (1.12) events with 2.07 (0.46) cosmic-ray-induced events, 1.19 (0.39) NC events and 0.51 (0.23) other beam backgrounds. The expected wrong sign contamination is of 7.24 $\bar{\nu}_\mu$ in $\nu_\mu$ beam and 12.58 $\nu_\mu$ in $\bar{\nu}_\mu$. FD data and best fit prediction can be seen in Fig.~\ref{FigFDnumu}. All above show a clear indication of both $\nu_\mu$ and $\bar{\nu}_\mu$ disappearance.
\section{Electron Neutrino and Antineutrino Appearance}
In order to maximize the statistical power of the $\nu_e$ selected events at FD, the sample is binned in both reconstructed energy and CVN score. There are two CVN bins of low and high purity (low and high PID), or ``core'' selection, and an additional ``peripheral'' bin. Events which fail containment or cosmic rejection cuts, but do have a very high CVN $\nu_e$ CC score, may be added to the peripheral sample. Because the events on the periphery are not always fully contained they are summed into a single bin instead of estimating their energy (up to reconstructed 4.5 GeV). The overall integrated selection efficiency of $\nu_e$ ($\bar{\nu}_e$) is 62\% (67\%), beam backgrounds are reduced by 95\% (99\%), the purity of the final predicted FD samples depends on the oscillation parameters, but ranges from 57\% (55\%) to 78\% (77\%).
\par
\begin{figure}
	\begin{center}
	\includegraphics[width=.496\linewidth]{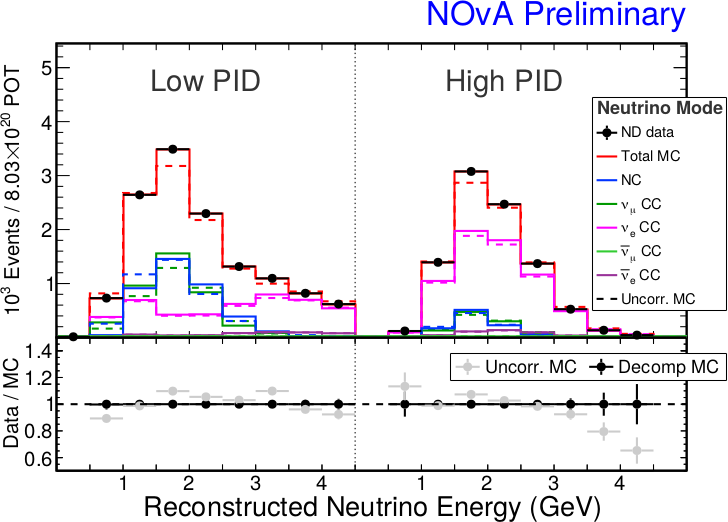}
	\includegraphics[width=.496\linewidth]{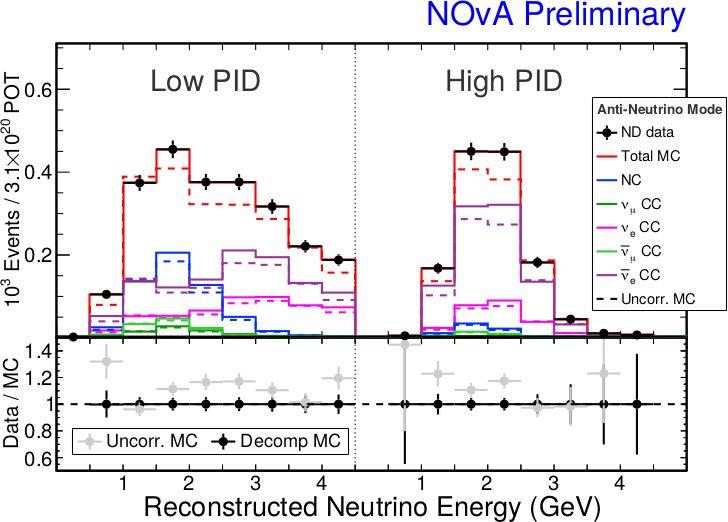}
	\caption{ND selected $\nu_e$ (left) and $\bar{\nu}_e$ (right) reconstructed energy data (black dots), uncorrected simulation (dashed red) and data-driven correction (solid red). The selection is decomposed (broken down) into NC (blue), $\nu_\mu/\bar{\nu}_\mu$ CC (dark/light green) and $\nu_e/\bar{\nu}_e$ CC (light/dark magenta). Binned in two PID bins, which are correlated to lower and higher purity of $\nu_e+\bar{\nu}_e$.}
	\label{FigNDnue}
	\end{center}
\end{figure}

 To estimate FD beam background F/N technique is used with ND $\nu_e$ sample. It consists of beam $\nu_e$ and $\nu_\mu$ CC or NC interactions misidentified as $\nu_e$ CC. Since each of these components oscillate differently along the way to the FD, the sample needs to be broken down into them. In the case of neutrino beam, $\nu_e$ component is constrained by inspecting the low-energy and high-energy $\nu_\mu$ CC spectra to adjust the yields of the parent hadrons that decay into both $\nu_\mu$ and $\nu_e$ (track $\nu_\mu$ and $\nu_e$ to their common parents). The $\nu_\mu$ component is estimated from observed distributions of time-delayed electrons from stopping $\mu$ decay. The rest is attributed to NC interaction. In the case of antineutrino beam, the components are only evenly and proportionally scaled to match ND data in each bin. ND selections and their breakdowns, or ``decomposition'', can be seen in Fig.~\ref{FigNDnue}. The high PID bin is dominated by the beam $\nu_e+\bar{\nu}_e$, the low PID bin has a significant admixture of $\nu_\mu$ ($\bar{\nu}_\mu$) CC and NC events. The beam background of FD peripheral bin is estimated from the high PID bin of the core sample. 
\par
\begin{figure}
	\begin{center}
	\includegraphics[width=.47\linewidth]{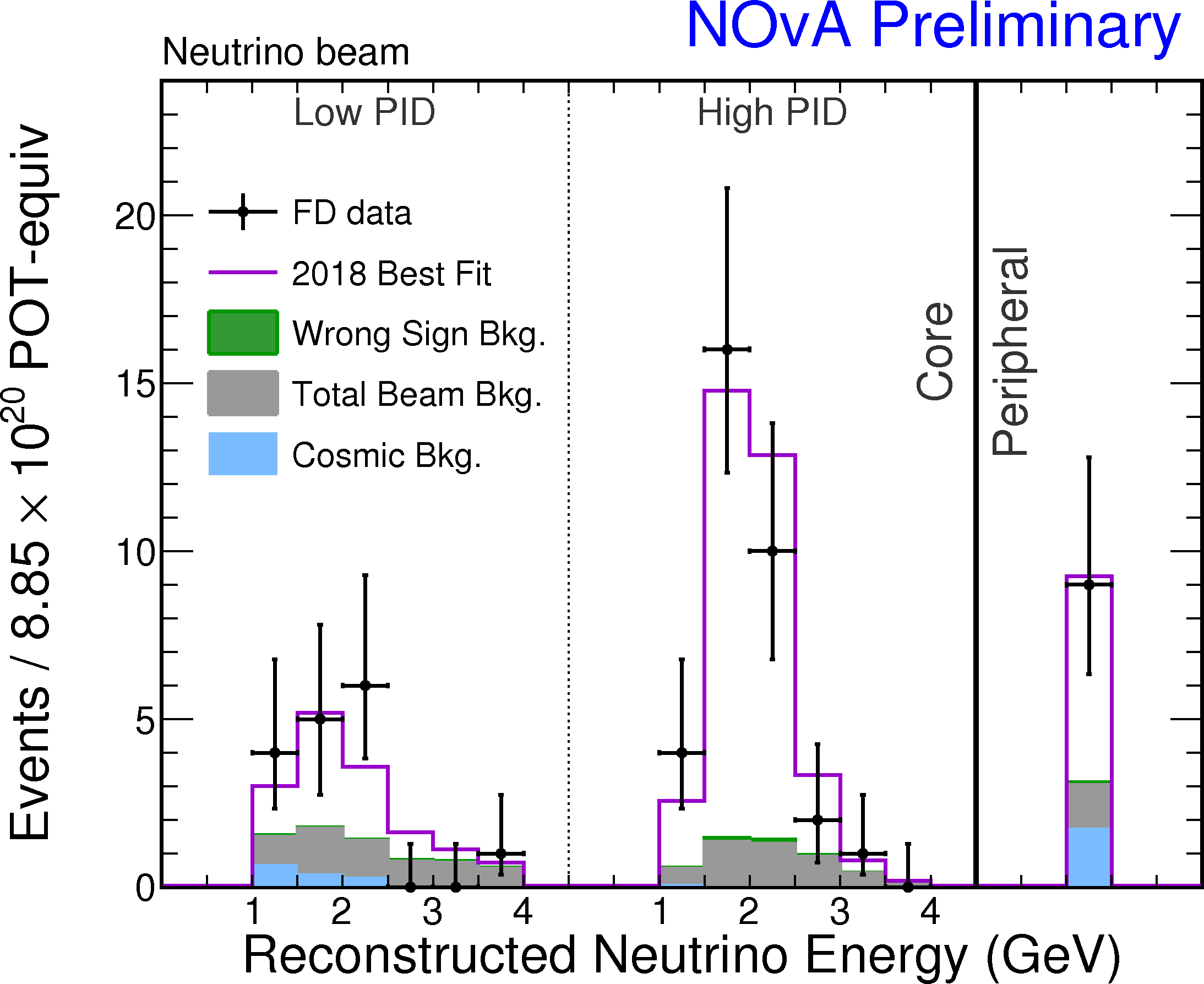}\hspace*{.05\linewidth}
	\includegraphics[width=.47\linewidth]{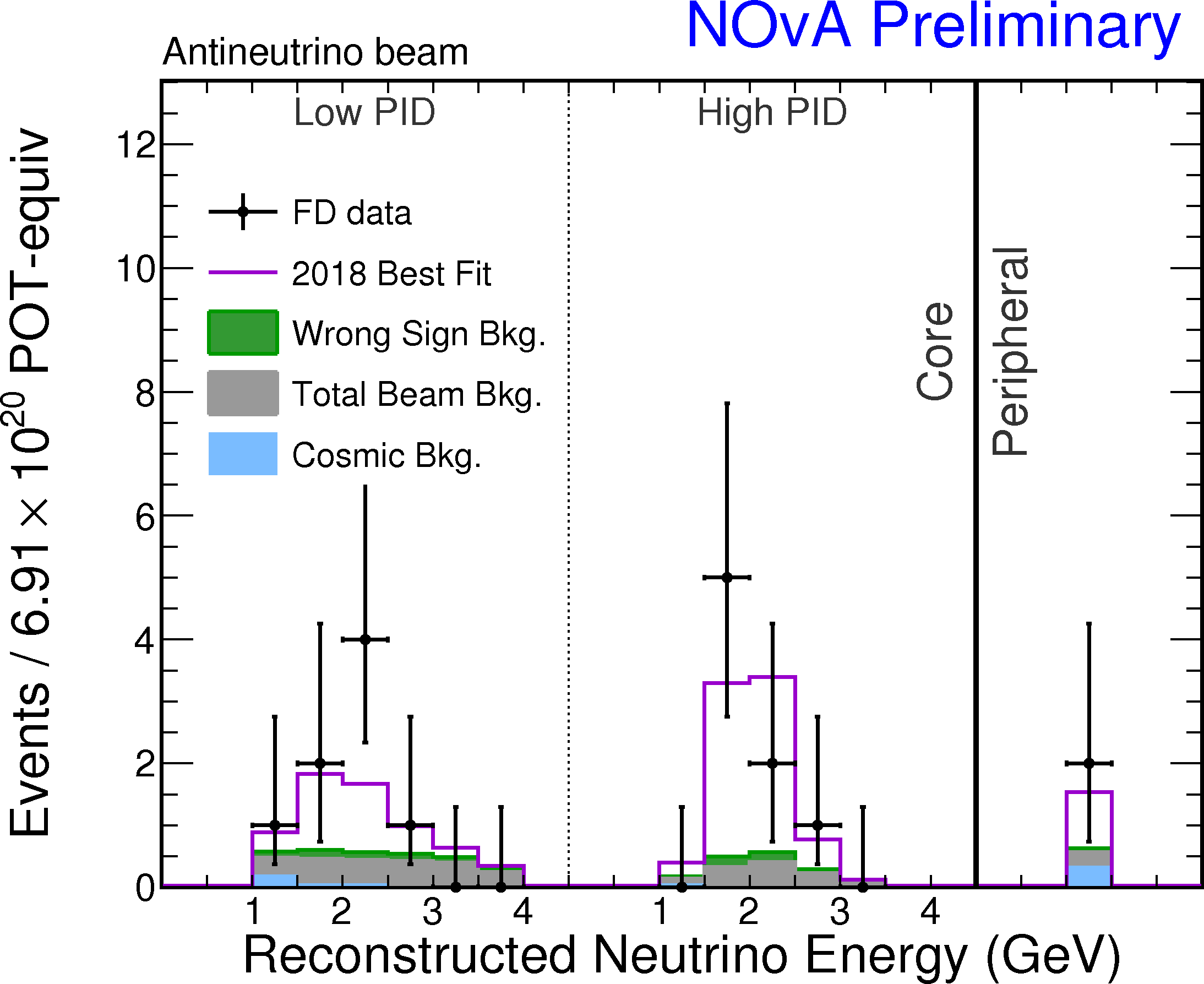}
	\end{center}
	\caption{FD data (black dots) selected $\nu_e$ (left) and $\bar{\nu}_e$ (right) candidates reconstructed energy binned in low and high PID bins and peripheral sample with energies up to 4.5 GeV. Best fit prediction (purple) shows the expected background of wrong sign (green), other beam background (grey) and cosmics (blue) as shaded areas.}
	\label{FigFDnue}
\end{figure}

 There are 58 (18) $\nu_e$ ($\bar{\nu}_e$) candidates in FD data with the prediction of 30 to~75 (10 to 22) depending on oscillation parameters ($\delta_\mathrm{CP}$, $\theta_{23}$ and NH or IH). The total expected background is 15.1 (5.3) events of 6.85 (2.57) beam $\nu_e+\bar{\nu}_e$, 0.63 (0.07) $\nu_\mu+\bar{\nu}_\mu$, 0.37 (0.15) $\nu_\tau+\bar{\nu}_\tau$, 3.21 (0.67) NC events, 3.33 (0.71) cosmic-ray-induced events and 0.66 $\bar{\nu}_e$ (2.57 $\nu_e$) from wrong sign component of the $\nu_\mu$ ($\bar{\nu}_\mu$) sample. The FD data and best fit predictions can be seen in Fig.~\ref{FigFDnue}. Antineutrino data gives more than 4$\sigma$ evidence of $\bar{\nu}_e$ appearance in $\bar{\nu}_\mu$ beam.
\section{Constraints on Oscillation Parameters}
To obtain oscillation parameters, a simultaneous fit of joint $\nu_e+\nu_\mu$ and both neutrino and antineutrino data was performed. Systematic uncertainties are incorporated as nuisance parameters with Gaussian penalty term, appropriately correlated between all the data sets. The leading systematics are worth a note: detector calibration (calorimetric energy scale), light production and collection model and muon energy scale (abs.+rel.) for $\nu_\mu$ disappearance; detector response and calibration, neutrino cross-sections and actual ND to FD differences for $\nu_e$ appearance. Several oscillation parameters are taken as inputs from other measurements: solar parameters $\theta_{12}$ and $\Delta m_{12}^2$, the mixing angle $\theta_{13}$ and its uncertainty was taken from reactor experiments, all in Ref.~\cite{PDG}. The best fit is
\begin{equation}
\Delta m_{32}^2 = 2.51^{+0.12}_{-0.08} \times 10^{-3}~\mathrm{eV}^2, \hspace*{5mm} \sin^2 \theta_{23} = 0.58\pm0.03, \hspace*{5mm} \delta_\mathrm{CP} = 0.17\pi,
\end{equation}
which corresponds to normal hierarchy and the uppper $\theta_{23}$ octant (UO, $\theta_{23} > 45^\circ$). All confidence levels (C.L.) and contours are constructed following the Feldman-Cousins approach \cite{FeldmanCousin}.
\par
\begin{figure}
	\begin{center}
	\includegraphics[width=.7\linewidth]{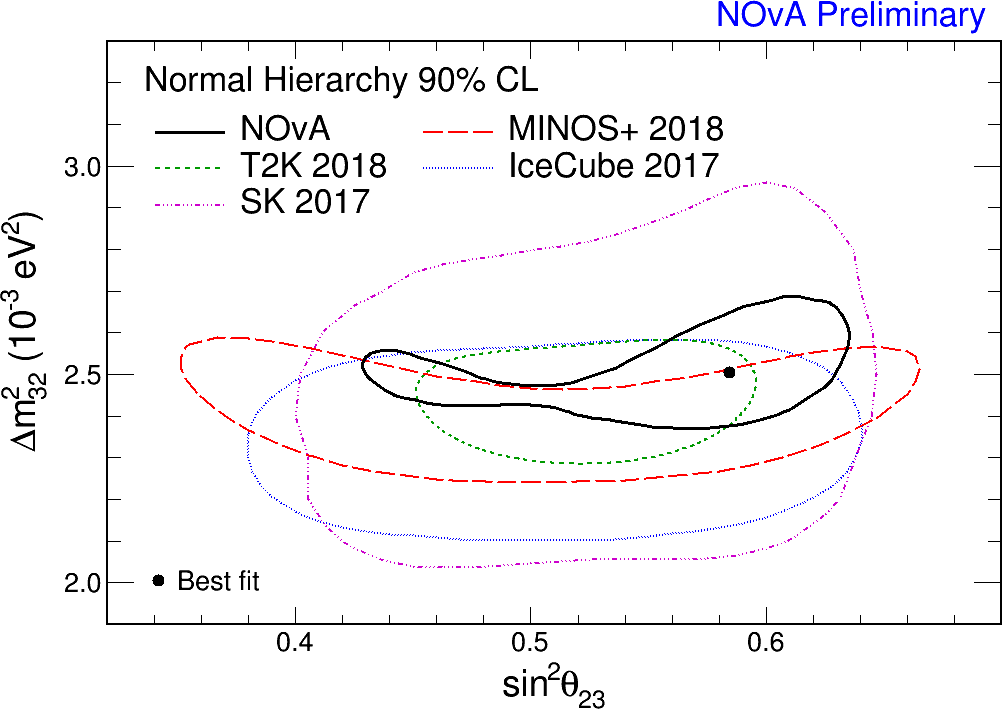}
	\end{center}
	\caption{Comparison of the allowed regions of $\Delta m_{32}^2$ vs. $\sin^2 \theta_{23}$ parameter space at 90\% C.L. as obtained by recent experiments: NOvA (black solid, black dot labels the best fit value), T2K (green dashed), MINOS+ (red dashed), IceCube (blue dotted) and Super-Kamiokande (purple dash-dotted).}
	\label{Fig23Conts}
\end{figure}

 The 90\% C.L. allowed region for a combination of $\Delta m_{32}^2$ versus $\sin^2 \theta_{23}$ in the $\Delta m_{32}^2 > 0$ half-plane, together with other results from MINOS+ (2018) \cite{MINOS+}, T2K (2017) \cite{T2K}, IceCube (2017) \cite{IceCube} and Super-Kamiokande (2017) \cite{Super-K} overlaid is shown in Fig.~\ref{Fig23Conts}. There is a clear consistency within all experiments despite that NOvA data asymetrically points to UO and rejects maximal 23 mixing ($\sin^2 \theta_{23} = 0.5$) at about 1.8$\sigma$ C.L.
\par
\begin{figure}
	\begin{center}
	\includegraphics[width=.496\linewidth]{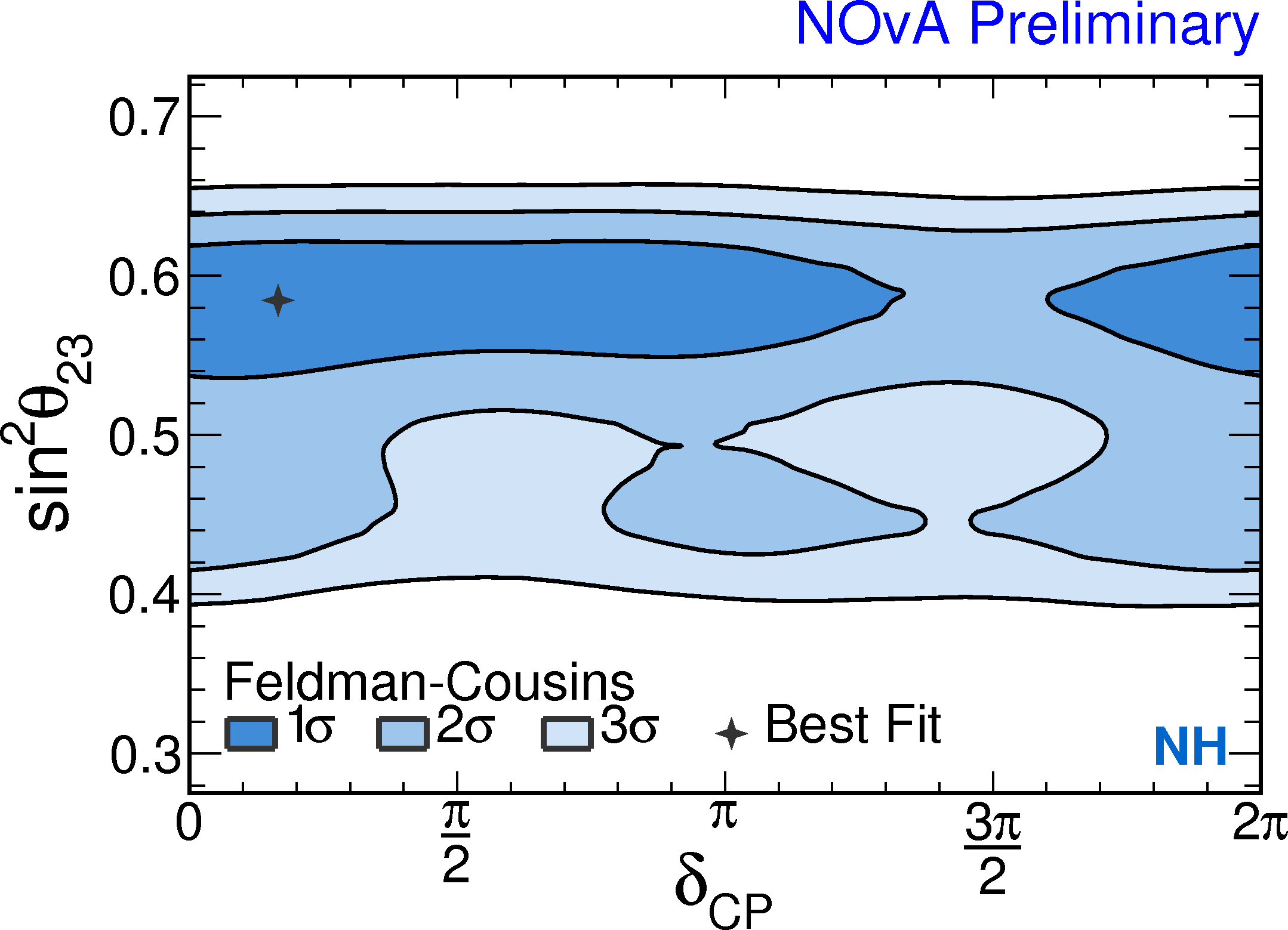}
	\includegraphics[width=.496\linewidth]{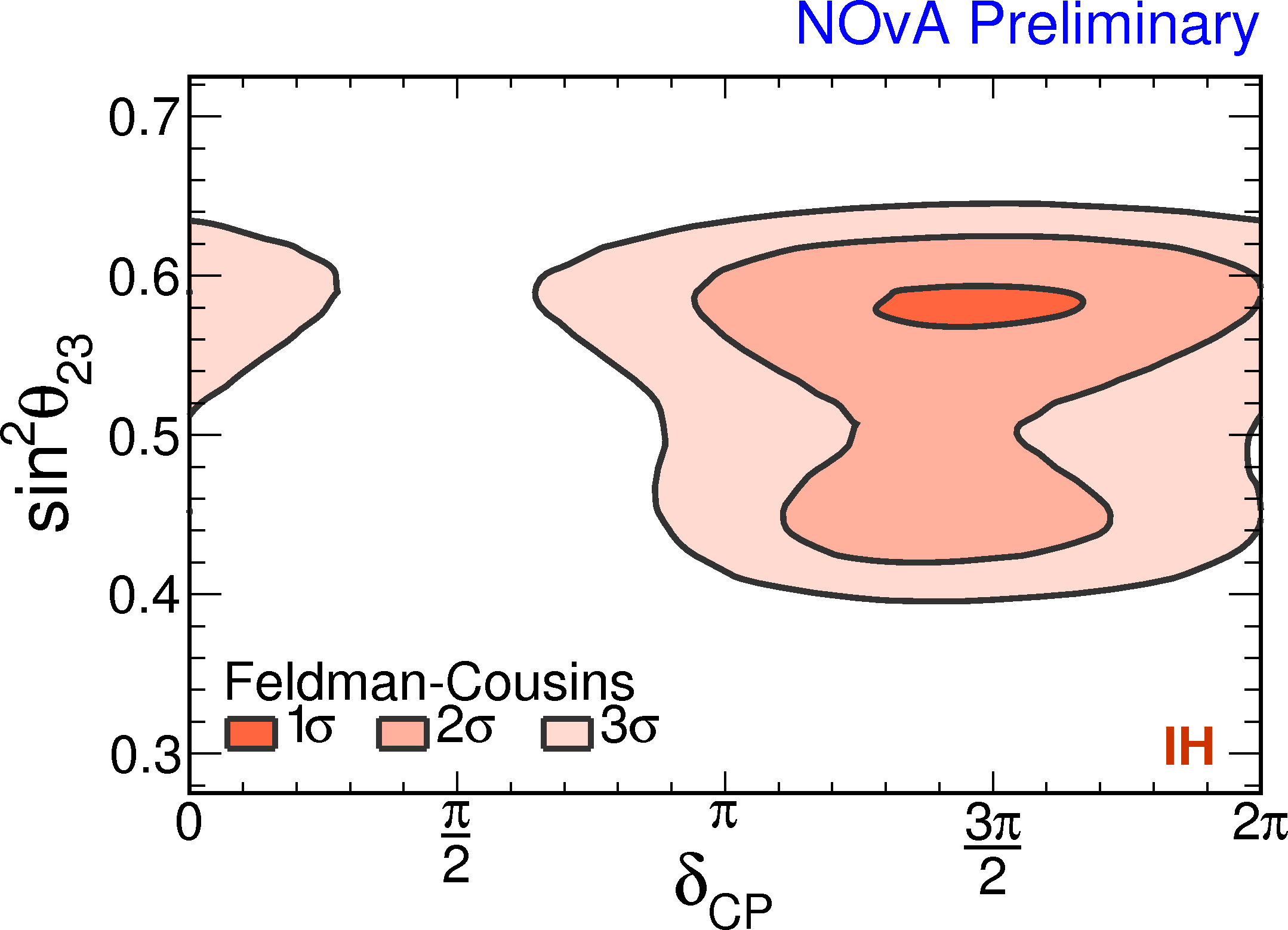}
	\end{center}
	\caption{1, 2 and 3$\sigma$ allowed regions of $\sin^2 \theta_{23}$ vs. $\delta_\mathrm{CP}$ neutrino oscillation parameter space consistent with $\nu_e$ appearance and $\nu_\mu$ disappearance data. The left plot corresponds to the case of normal hierarchy of neutrino masses ($\Delta m_{32}^2 > 0$), the left one to the inverted hierarchy ($\Delta m_{32}^2 < 0$).}
	\label{FigDeltaConts}
\end{figure}

 Fig.~\ref{FigDeltaConts} shows the 1, 2 and 3$\sigma$ C.L. allowed regions for $\sin^2 \theta_{23}$ versus $\delta_\mathrm{CP}$ in both cases of NH and IH (mass ordering). It is worth noticing, that the values of $\delta_\mathrm{CP}$ around $\pi/2$ are excluded at $> 3\sigma$ C.L. for IH, similarly to previous NOvA neutrino only analysis \cite{NOvA2018paper}. On the other hand, rather weak constraints on $\delta_\mathrm{CP}$ itself allow all possible values [0,2$\pi$] in 2$\sigma$ interval for the case of NH and UO.
\section{Future Prospects}
NOvA is expected to run until 2024 with about an equal total exposure of neutrino and antineutrino beam. Moreover, several accelerator upgrades to enhance the beam performance are planned for the next years. Based on these prerequisities and projected 2018 analysis techniques there is a possibility of more than 3$\sigma$ sensitivity to hierarchy resolution by 2020 in case of favorable true values of oscillation parameters (NH and $\delta_\mathrm{CP} =3\pi/2$), or by 2024 for 30-50\% of all possible $\delta_\mathrm{CP}$ otherwise. Besides that, about 3$\sigma$ sensitivity to $\theta_{23}$ octant determination and more than $2\sigma$ to CP violation in case of $\delta_\mathrm{CP}=\pi/2$ or $3\pi/2$ (maximal violation) are expected by 2024.
\par To further improve neutrino oscillation analysis and to extend the reach of the experiment, NOvA plans to start an intensive test beam program in early 2019. The main focus will be on simulation tuning, systematics study and their reduction, validation and training of reconstruction or machine learning algorithms.
\section{Summary}
The first antineutrino data from NOvA (6.91$\times 10^{20}$ POT) has been analyzed together with existing neutrino data (8.85$\times 10^{20}$ POT). The measurements are well consistent with the standard oscillation model of 3 active neutrino flavors. NOvA observes more than 4$\sigma$ evidence for $\bar{\nu}_e$ appearance in $\bar{\nu}_\mu$ beam. The results of joint analysis of neutrino and antineutrino and both $\nu_\mu$ disappearance and $\nu_e$ appearance channels give the parameters estimates of $\sin^2 \theta_{23} =0.58 \pm 0.03$ and $\Delta m_{32}^2 = 2.51^{+0.12}_{-0.08} \times 10^{-3}$~eV$^2$, which are in a good agreement with other accelerator and atmospheric oscillation experiments. There is an indication of non-maximal 23 mixing at 1.8$\sigma$ and inkling of upper octant of $\theta_{23}$ angle. Data also prefers normal hierarchy of neutrino masses at 1.8$\sigma$, while simultaneously disfavoring inverted hierarchy for $\delta_\mathrm{CP}$ around $3\pi/2$ at more than 3$\sigma$. NOvA plans to continue running until 2024 in both neutrino and antineutrino beam modes.

\paragraph*{Acknowledgments} I would like to thank the organizers for a great conference of NTIHEP 2018. This work is supported by MSMT CR (Ministry of Education, Youth and Sports, Czech Republic).

\end{document}